\documentclass[10pt,a4paper,onecolumn]{article}
\usepackage{marginnote}
\usepackage{graphicx}
\usepackage{xcolor}
\usepackage{authblk,etoolbox}
\usepackage{titlesec}
\usepackage{calc}
\usepackage{tikz}
\usepackage{hyperref}
\hypersetup{colorlinks,breaklinks,
            urlcolor=[rgb]{0.0, 0.5, 1.0},
            linkcolor=[rgb]{0.0, 0.5, 1.0}}
\usepackage{caption}
\usepackage{tcolorbox}
\usepackage{amssymb,amsmath}
\usepackage{ifxetex,ifluatex}
\usepackage{seqsplit}
\usepackage{xstring}

\usepackage{float}
\let\origfigure\figure
\let\endorigfigure\endfigure
\renewenvironment{figure}[1][2] {
    \expandafter\origfigure\expandafter[H]
} {
    \endorigfigure
}

\usepackage{fixltx2e} 
\usepackage[
  backend=biber,
]{biblatex}
\bibliography{paper.bib}

\let\textttOrig=\texttt
\def\texttt#1{\expandafter\textttOrig{\seqsplit{#1}}}
\renewcommand{\seqinsert}{\ifmmode
  \allowbreak
  \else\penalty6000\hspace{0pt plus 0.02em}\fi}

\makeatletter
\let\href@Orig=\href
\def\href@Urllike#1#2{\href@Orig{#1}{\begingroup
    \def\Url@String{#2}\Url@FormatString
    \endgroup}}
\def\href@Notdoi#1#2{\def\tempa{#1}\def\tempb{#2}%
  \ifx\tempa\tempb\relax\href@Urllike{#1}{#2}\else
  \href@Orig{#1}{#2}\fi}
\def\href#1#2{%
  \IfBeginWith{#1}{https://doi.org}%
  {\href@Urllike{#1}{#2}}{\href@Notdoi{#1}{#2}}}
\makeatother

\usepackage[top=3.5cm, bottom=3cm, right=1.5cm, left=1.0cm,
            headheight=2.2cm, reversemp, includemp, marginparwidth=4.5cm]{geometry}

\titleformat{\section}
  {\normalfont\sffamily\Large\bfseries}
  {}{0pt}{}
\titleformat{\subsection}
  {\normalfont\sffamily\large\bfseries}
  {}{0pt}{}
\titleformat{\subsubsection}
  {\normalfont\sffamily\bfseries}
  {}{0pt}{}
\titleformat*{\paragraph}
  {\sffamily\normalsize}

\usepackage{fancyhdr}
\makeatletter
\let\ps@plain\ps@fancy
\fancyheadoffset[L]{4.5cm}
\fancyfootoffset[L]{4.5cm}

\definecolor{linky}{rgb}{0.0, 0.5, 1.0}

\newtcolorbox{repobox}
   {colback=red, colframe=red!75!black,
     boxrule=0.5pt, arc=2pt, left=6pt, right=6pt, top=3pt, bottom=3pt}

\newcommand{\ExternalLink}{%
   \tikz[x=1.2ex, y=1.2ex, baseline=-0.05ex]{%
       \begin{scope}[x=1ex, y=1ex]
           \clip (-0.1,-0.1)
               --++ (-0, 1.2)
               --++ (0.6, 0)
               --++ (0, -0.6)
               --++ (0.6, 0)
               --++ (0, -1);
           \path[draw,
               line width = 0.5,
               rounded corners=0.5]
               (0,0) rectangle (1,1);
       \end{scope}
       \path[draw, line width = 0.5] (0.5, 0.5)
           -- (1, 1);
       \path[draw, line width = 0.5] (0.6, 1)
           -- (1, 1) -- (1, 0.6);
       }
   }

\patchcmd{\@maketitle}{center}{flushleft}{}{}
\patchcmd{\@maketitle}{center}{flushleft}{}{}
\patchcmd{\@maketitle}{\LARGE}{\LARGE\sffamily}{}{}
\def\maketitle{{%
  
  \AB@maketitle}}
\makeatletter
\renewcommand\AB@affilsepx{ \protect\Affilfont}
\renewcommand\AB@affilnote[1]{{\bfseries #1}\hspace{3pt}}
\renewcommand{\affil}[2][]%
   {\newaffiltrue\let\AB@blk@and\AB@pand
      \if\relax#1\relax\def\AB@note{\AB@thenote}\else\def\AB@note{#1}%
        \setcounter{Maxaffil}{0}\fi
        \begingroup
        \let\href=\href@Orig
        \let\texttt=\textttOrig
        \let\protect\@unexpandable@protect
        \def\thanks{\protect\thanks}\def\footnote{\protect\footnote}%
        \@temptokena=\expandafter{\AB@authors}%
        {\def\\{\protect\\\protect\Affilfont}\xdef\AB@temp{#2}}%
         \xdef\AB@authors{\the\@temptokena\AB@las\AB@au@str
         \protect\\[\affilsep]\protect\Affilfont\AB@temp}%
         \gdef\AB@las{}\gdef\AB@au@str{}%
        {\def\\{, \ignorespaces}\xdef\AB@temp{#2}}%
        \@temptokena=\expandafter{\AB@affillist}%
        \xdef\AB@affillist{\the\@temptokena \AB@affilsep
          \AB@affilnote{\AB@note}\protect\Affilfont\AB@temp}%
      \endgroup
       \let\AB@affilsep\AB@affilsepx
}
\makeatother

\renewcommand\Affilfont{\sffamily\small\mdseries}
\ifnum 0\ifxetex 1\fi\ifluatex 1\fi=0 
  \usepackage[T1]{fontenc}
  \usepackage[utf8]{inputenc}

\else 
  \ifxetex
    \usepackage{mathspec}
  \else
    \usepackage{fontspec}
  \fi
  \defaultfontfeatures{Ligatures=TeX,Scale=MatchLowercase}

\fi
\IfFileExists{upquote.sty}{\usepackage{upquote}}{}
\IfFileExists{microtype.sty}{%
\usepackage{microtype}
\UseMicrotypeSet[protrusion]{basicmath} 
}{}

\usepackage{hyperref}
\hypersetup{unicode=true,
            pdftitle={Your: Your Unified Reader},
            pdfborder={0 0 0},
            breaklinks=true}
\let\addcontentslineOrig=\addcontentsline
\def\addcontentsline#1#2#3{\bgroup
  \let\texttt=\textttOrig\addcontentslineOrig{#1}{#2}{#3}\egroup}
\let\markbothOrig\markboth
\def\markboth#1#2{\bgroup
  \let\texttt=\textttOrig\markbothOrig{#1}{#2}\egroup}
\let\markrightOrig\markright
\def\markright#1{\bgroup
  \let\texttt=\textttOrig\markrightOrig{#1}\egroup}

\usepackage{graphicx,grffile}
\makeatletter
\def\maxwidth{\ifdim\Gin@nat@width>\linewidth\linewidth\else\Gin@nat@width\fi}
\def\maxheight{\ifdim\Gin@nat@height>\textheight\textheight\else\Gin@nat@height\fi}
\makeatother
\setkeys{Gin}{width=\maxwidth,height=\maxheight,keepaspectratio}
\IfFileExists{parskip.sty}{%
\usepackage{parskip}
}{
\setlength{\parindent}{0pt}
\setlength{\parskip}{6pt plus 2pt minus 1pt}
}
\ifx\paragraph\undefined\else
\let\oldparagraph\paragraph
\renewcommand{\paragraph}[1]{\oldparagraph{#1}\mbox{}}
\fi
\ifx\subparagraph\undefined\else
\let\oldsubparagraph\subparagraph
\renewcommand{\subparagraph}[1]{\oldsubparagraph{#1}\mbox{}}
\fi

\title{Your: Your Unified Reader}

        \author[1, 2]{Kshitij Aggarwal}
          \author[1, 2]{Devansh Agarwal}
          \author[1, 2]{Joseph W Kania}
          \author[1, 2]{William Fiore}
          \author[1, 2]{Reshma Anna Thomas}
          \author[3]{Scott M. Ransom}
          \author[4]{Paul B. Demorest}
          \author[5]{Robert S. Wharton}
          \author[1, 2]{Sarah Burke-Spolaor}
          \author[1, 2]{Duncan R. Lorimer}
          \author[1, 2]{Maura A. Mclaughlin}
          \author[1, 2]{Nathaniel Garver-Daniels}
    
      \affil[1]{West Virginia University, Department of Physics and Astronomy, P. O. Box
6315, Morgantown 26506, WV, USA}
      \affil[2]{Center for Gravitational Waves and Cosmology, West Virginia University,
Chestnut Ridge Research Building, Morgantown 26506, WV, USA}
      \affil[3]{National Radio Astronomy Observatory, Charlottesville, VA 22903, USA}
      \affil[4]{National Radio Astronomy Observatory, Socorro, NM, 87801, USA}
      \affil[5]{Max-Planck-Institut für Radioastronomie, Auf dem Hügel 69, D-53121 Bonn,
Germany}
  \date{\vspace{-5ex}}

\begin{document}
\maketitle

\marginpar{
  \sffamily\small

  {\bfseries DOI:} \href{https://doi.org/10.21105/joss.02750}{\color{linky}{10.21105/joss.02750}}

  \vspace{2mm}

  {\bfseries Software}
  \begin{itemize}
    \setlength\itemsep{0em}
    \item \href{https://github.com/openjournals/joss-reviews/issues/2750}{\color{linky}{Review}} \ExternalLink
    \item \href{https://github.com/thepetabyteproject/your}{\color{linky}{Repository}} \ExternalLink
    \item \href{https://doi.org/10.5281/zenodo.4269947}{\color{linky}{Archive}} \ExternalLink
  \end{itemize}

  \vspace{2mm}

  {\bfseries Submitted:} 28 August 2020\\
  {\bfseries Published:} 15 November 2020

  \vspace{2mm}
  {\bfseries License}\\
  Authors of papers retain copyright and release the work under a Creative Commons Attribution 4.0 International License (\href{https://creativecommons.org/licenses/by/4.0/}{\color{linky}{CC BY 4.0}}).
}

\hypertarget{summary}{%
\section{Summary}\label{summary}}

The understanding of fast radio transients like pulsar single pulses,
rotating radio transients (RRATs), and especially Fast Radio Bursts
(FRBs) has evolved rapidly over the last decade. This is primarily due
to dedicated campaigns by sensitive radio telescopes to search for
transients. The advancement in signal processing and GPU processing
systems has enabled new transient detectors at various telescopes to
perform much more sensitive searches than their predecessors due to the
ability to find and process FRB candidates in real-time or
near-real-time. Typically the data output from the telescopes is in one
of the two commonly used formats: psrfits (Hotan, van Straten, and
Manchester 2004) and
\href{http://sigproc.sourceforge.net/}{\texttt{Sigproc\ filterbank}}
(Lorimer 2011). Software developed for transient searches often only
works with one of these two formats, limiting their general
applicability. Therefore, researchers have to write custom scripts to
read/write the data in their format of choice before they can begin any
data analysis relevant for their research. This has led to the
development of several python libraries to manage one or the other data
format (like \href{https://github.com/demorest/pysigproc}{pysigproc},
\href{https://github.com/scottransom/presto/blob/master/python/presto/psrfits.py}{psrfits},
\href{https://github.com/FRBs/sigpyproc3}{sigpyproc}, etc). Still, no
general tool exists which can work across data formats.

\hypertarget{statement-of-need}{%
\section{Statement of need}\label{statement-of-need}}

\texttt{Your} (Your Unified Reader) is a python-based library that
unifies the data processing across multiple commonly used formats.
\texttt{Your} was conceived initially to perform data ingestion for The
Petabyte FRB search Project (TPP), which will uniformly search a large
number of datasets from telescopes around the world for FRBs. As this
project is going to process data in different formats from multiple
telescopes worldwide, a unified reader was required to streamline the
search pipeline. \texttt{Your} implements a user-friendly interface to
read and write in the data format of choice. It also generates unified
metadata corresponding to the input data file for a quick understanding
of observation parameters and provides utilities to perform common data
analysis operations. \texttt{Your} also provides several
state-of-the-art radio frequency interference mitigation (RFI)
algorithms (Agarwal, Lorimer, et al. 2020; Nita and Gary 2010), which
can now be used during any stage of data processing (reading, writing,
etc.) to filter out artificial signals.

\texttt{Your} can be used at the data ingestion step of any transient
search pipeline and can provide data and observation parameters in a
format-independent manner. Generic tools can thus be used to perform the
search and further data analysis. It also enables online processing like
RFI flagging, decimation, subband search, etc.; functions for some of
these are already available in \texttt{Your}. It can also be used to
perform analysis of individual candidate events (using
\texttt{Candidate} class): generate candidate data cutouts, create
publication-ready visualizations, and perform GPU accelerated
pre-processing for candidate classification (Agarwal, Aggarwal, et al.
2020). It also consists of functions to run commonly used single-pulse
search software
\href{https://sourceforge.net/projects/heimdall-astro/}{\texttt{Heimdall}}
(Barsdell 2012) on any input data format.

\texttt{Your} will not only benefit experienced researchers but also new
undergraduate and graduate students who otherwise have to face a
significant bottleneck to understand various data formats and develop
custom tools to access the data before any analysis can be done on it.
Moreover, \texttt{Your} is written purely in python, which is a commonly
used language within Astronomy. It also comes with comprehensive
\href{https://thepetabyteproject.github.io/your/}{documentation} and
\href{https://github.com/thepetabyteproject/your/tree/master/examples}{example
notebooks} to make it easier to get started.

\texttt{Your} uses the matplotlib library (Hunter 2007) for plotting,
and also makes use of various numpy (Harris et al. 2020), scipy
(Virtanen et al. 2020), scikit-image (Van der Walt et al. 2014), numba
(Lam, Pitrou, and Seibert 2015) and Pandas (The pandas development team
2020; McKinney 2010) functions. \texttt{Your} also leverages several
functions in the Astropy package (Astropy Collaboration et al. 2013;
Price-Whelan et al. 2018): fits (astropy.io.fits), units
(astropy.units), coordinates (astropy.coordinates) and time
(astropy.time).

\hypertarget{acknowledgements}{%
\section{Acknowledgements}\label{acknowledgements}}

KA, DA, WF, SMR, PDB, SBS, DRL, MAM, and NGD are members of the NANOGrav
Physics Frontiers Center, supported by NSF award number 1430284. MAM,
DA, DRL, JWK, and SBS are also supported by NSF award number 1458952.
DRL, MAM, DA and JWK acknowledge support from the NSF award AAG-1616042.
WF acknowledges funding from the WVU STEM Mountains of Excellence
graduate fellowship. RSW acknowledges financial support by the European
Research Council (ERC) for the ERC Synergy Grant BlackHoleCam under
contract no. 610058.

\hypertarget{references}{%
\section*{References}\label{references}}
\addcontentsline{toc}{section}{References}

\hypertarget{refs}{}
\leavevmode\hypertarget{ref-fetch2020}{}%
Agarwal, Devansh, Kshitij Aggarwal, Sarah Burke-Spolaor, Duncan R.
Lorimer, and Nathaniel Garver-Daniels. 2020. ``FETCH: A deep-learning
based classifier for fast transient classification.'' \emph{Monthly
Notices of the Royal Astronomical Society} 497 (2): 1661--74.
\url{https://doi.org/10.1093/mnras/staa1856}.

\leavevmode\hypertarget{ref-agarwal2020}{}%
Agarwal, Devansh, D. R. Lorimer, M. P. Surnis, X. Pei, A. Karastergiou,
G. Golpayegani, D. Werthimer, et al. 2020. ``Initial results from a
real-time FRB search with the GBT.'' \emph{Monthly Notices of the Royal
Astronomical Society} 497 (1): 352--60.
\url{https://doi.org/10.1093/mnras/staa1927}.

\leavevmode\hypertarget{ref-astropy:2013}{}%
Astropy Collaboration, T. P. Robitaille, E. J. Tollerud, P. Greenfield,
M. Droettboom, E. Bray, T. Aldcroft, et al. 2013. ``Astropy: A community
Python package for astronomy.'' \emph{Astronomy \& Astrophysics} 558
(October): A33. \url{https://doi.org/10.1051/0004-6361/201322068}.

\leavevmode\hypertarget{ref-barsdell2012heimdall}{}%
Barsdell, B. R. 2012. ``Advanced architectures for astrophysical
supercomputing.'' PhD thesis, Swinburne University of Technology.

\leavevmode\hypertarget{ref-harris2020}{}%
Harris, Charles R., K. Jarrod Millman, Stéfan J. van der Walt, Ralf
Gommers, Pauli Virtanen, David Cournapeau, Eric Wieser, et al. 2020.
``Array Programming with NumPy.'' \emph{Nature} 585 (7825). Springer
Science; Business Media LLC: 357--62.
\url{https://doi.org/10.1038/s41586-020-2649-2}.

\leavevmode\hypertarget{ref-hotan2004}{}%
Hotan, A. W., W. van Straten, and R. N. Manchester. 2004. ``PSRCHIVE and
PSRFITS: An Open Approach to Radio Pulsar Data Storage and Analysis.''
\emph{Publications of the Astronomical Society of Australia} 21 (3):
302--9. \url{https://doi.org/10.1071/AS04022}.

\leavevmode\hypertarget{ref-Hunter:2007}{}%
Hunter, J. D. 2007. ``Matplotlib: A 2D graphics environment.''
\emph{Computing in Science \& Engineering} 9 (3). IEEE COMPUTER SOC:
90--95. \url{https://doi.org/10.1109/MCSE.2007.55}.

\leavevmode\hypertarget{ref-numba}{}%
Lam, Siu Kwan, Antoine Pitrou, and Stanley Seibert. 2015. ``Numba: A
LLVM-Based Python JIT Compiler.'' In \emph{Proceedings of the Second
Workshop on the Llvm Compiler Infrastructure in Hpc}. LLVM '15. New
York, NY, USA: Association for Computing Machinery.
\url{https://doi.org/10.1145/2833157.2833162}.

\leavevmode\hypertarget{ref-sigproc}{}%
Lorimer, D. R. 2011. ``SIGPROC: Pulsar Signal Processing Programs.''

\leavevmode\hypertarget{ref-pandas2010}{}%
McKinney. 2010. ``Data Structures for Statistical Computing in Python.''
In \emph{Proceedings of the 9th Python in Science Conference}, edited by
Stéfan van der Walt and Jarrod Millman, 56--61.
\href{https://doi.org/\%2010.25080/Majora-92bf1922-00a\%20}{https://doi.org/ 10.25080/Majora-92bf1922-00a}.

\leavevmode\hypertarget{ref-nita2010}{}%
Nita, G. M., and D. E. Gary. 2010. ``The generalized spectral kurtosis
estimator.'' \emph{Monthly Notices of the Royal Astronomical Society}
406 (1): L60--L64.
\url{https://doi.org/10.1111/j.1745-3933.2010.00882.x}.

\leavevmode\hypertarget{ref-astropy:2018}{}%
Price-Whelan, A. M., B. M. Sipőcz, H. M. Günther, P. L. Lim, S. M.
Crawford, S. Conseil, D. L. Shupe, et al. 2018. ``The Astropy Project:
Building an Open-science Project and Status of the v2.0 Core Package.''
\emph{The Astronomical Journal} 156 (September): 123.
\url{https://doi.org/10.3847/1538-3881/aabc4f}.

\leavevmode\hypertarget{ref-reback2020pandas}{}%
The pandas development team. 2020. \emph{pandas-dev/pandas: Pandas}
(version latest). Zenodo. \url{https://doi.org/10.5281/zenodo.3509134}.

\leavevmode\hypertarget{ref-van2014scikit}{}%
Van der Walt, Stefan, Johannes L Schönberger, Juan Nunez-Iglesias,
François Boulogne, Joshua D Warner, Neil Yager, Emmanuelle Gouillart,
and Tony Yu. 2014. ``scikit-image: image processing in Python.''
\emph{PeerJ} 2. PeerJ Inc.: e453.

\leavevmode\hypertarget{ref-2020SciPy}{}%
Virtanen, Pauli, Ralf Gommers, Travis E. Oliphant, Matt Haberland, Tyler
Reddy, David Cournapeau, Evgeni Burovski, et al. 2020. ``SciPy 1.0:
Fundamental Algorithms for Scientific Computing in Python.''
\emph{Nature Methods} 17: 261--72.
\url{https://doi.org/10.1038/s41592-019-0686-2}.

\end{document}